\documentclass[aps,prb,twocolumn,groupedaddress,showpacs]{revtex4}
\usepackage{epsfig}
\usepackage{color}

\begin{document}

\title{Exact Numerical Study of Pair Formation with Imbalanced Fermion
Populations} 

\author{G. G. Batrouni$^1$, M. H. Huntley$^2$, V. G. Rousseau$^3$ and
R. T. Scalettar$^4$}

\affiliation{$^1$INLN, Universit\'e de Nice-Sophia Antipolis, CNRS; 
1361 route des Lucioles, 06560 Valbonne, France}

\affiliation{$^2$Physics Department, Massachusetts Insititute of
Technology, Cambridge MA 02139}

\affiliation{$^3$Lorentz Insitute, Leiden University, P.O.~Box 9506,
2300 RA, Leiden, The Netherlands}

\affiliation{$^4$Physics Department, University of California, Davis,
California 95616, USA}

\begin{abstract}
We present an exact Quantum Monte Carlo study of the attractive
1-dimensional Hubbard model with imbalanced fermion population.  The
pair-pair correlation function, which decays monotonically in the
absence of polarization $P$, develops oscillations when $P$ is
nonzero, characteristic of Fulde-Ferrell-Larkin-Ovchinnikov phase.
The pair momentum distribution peaks at a momentum equal to the
difference in the Fermi momenta.  At strong coupling, the minority and
majority momentum distributions are shown to be deformed, reflecting
the presence of the other species, and its Fermi surface.  The FFLO
oscillations survive the presence of a confining potential, and the
local polarization at the trap center exhibits a marked dip, similar
to that observed experimentally.
\end{abstract}

\maketitle

Following the observation of transitions into Mott insulating phases
as the ratio of interaction strength to kinetic energy is
varied\cite{greiner02}, ultracold atoms trapped on optical lattices
have been increasingly used to emulate the rich physics of strongly
correlated condensed matter systems.  One of the subtle phenomena
being sought is the Fulde-Ferrell-Larkin-Ovchinnikov (FFLO) phase
\cite{fulde64,larkin64}, in which an imbalance in the populations
of up and down spin electrons in a superconductor leads to
Bose-Einstein condensation in states at non-zero momentum and to
spatial inhomogeneities of the pairing and spin correlations.  The
observation of the FFLO phase in solids proved very difficult and was
only achieved recently in heavy fermion systems \cite{radovan03}.

Cold atom experiments, in which two hyperfine states of fermionic
atoms play the role of up and down spins, have now reported the
presence of pairing in the case of unequal populations
\cite{zwierlein06,partridge06}. Many theoretical studies using mean
field
\cite{castorina05,sheehy06,kinnunen06,machida06,gubbels06,parish07,hu07,he07,wilczek,koponen06},
effective Lagrangian \cite{son06} and Bethe ansatz \cite{orso07} have
been performed for the uniform system with extensions to the trapped
system using the local density approximation (LDA). For the uniform
case, the debate revolves on the details for the paired state: FFLO
pairs forming with non-zero momentum {\it vs} breached pairing (BP) at
zero momentum\cite{koponen06,wilczek,sarma}, and on the fragility of
such phases. No general consensus has emerged.

In this paper we report an exact Quantum Monte Carlo (QMC) study of
pairing in one dimensional fermion systems with population imbalance.
We focus first on the homogeneous (untrapped) case since even here
there is no agreement on the pairing mode. Our key result is that the
FFLO phase is very robust and appears to be the dominant pairing
mechanism.  We show that the pair Green function exhibits oscillations
characteristic of this phase, leading the pair momentum distribution
function to peak at a momentum corresponding to the difference of the
Fermi momenta of the individual species.

In the last section, we consider the trapped system and how the basic
FFLO phenomena survive the resulting density and polarization
inhomogeneity.  In addition, we show that at large $|U|$ the
difference in density between the two species exhibits a deep minimum
in the trap center: Not only is the minority species attracted to the
trap center by the majority potential, but in fact the two populations
attempt to equalize there.  This is a key signature of the experiments
on unequal populations\cite{zwierlein06,partridge06}.

We first describe the model and the QMC method we employed.  The
attractive Hubbard Hamiltonian is,
\begin{eqnarray}
  \label{Hamiltonian}
\hat\mathcal H=&-&t\sum_{l \sigma} (c_{l+1\,\sigma}^{\dagger}
       c_{l\,\sigma}^{\phantom\dagger} + c_{l\,\sigma}^\dagger c_{l+1
       \, \sigma}^{\phantom\dagger}) \nonumber \\ &+&U \sum_{l}
       n_{l\,1} n_{l\,2} + V_T \sum_{l} l^2 (n_{l \, 1} + n_{l \, 2}),
\end{eqnarray}
where $c_{j\,\sigma}^\dagger(c_{j\, \sigma})$ are fermion creation
(destruction) operators on spatial site $j$ with the fermionic species
labeled by $\sigma=1,2$ and $n_{j\,\sigma}=c_{j\,\sigma}^\dagger
c_{j\, \sigma}$ the corresponding number operator.  We take the
hopping parameter $t=1$ to set the energy scale and attractive on-site
interactions $U<0$.  $V_T$ is the strength of the quadratic confining
potential.  All our results are for inverse temperature $\beta=64$, so
that $T=W/256$ with $W=4t$ the bandwidth. We have verified that this
yields ground state properties.  We study a one dimensional lattice
with $L=32$ sites, unless otherwise stated.

For our simulations, we use a continuous imaginary time canonical
``worm" algorithm where the total number of particles is maintained
strictly constant \cite{pollet}. In this algorithm, two worms are
propagated, one for each type of fermion, which allows the calculation
of the real-space Green functions of the two fermionic species,
$G_\sigma$, and also the pair Green function, $G_{\rm pair}$, which are
defined by,
\begin{eqnarray}
G_\sigma(l) &=& \langle c_{j+l \,\sigma}^{\phantom\dagger} c_{j
\,\sigma}^\dagger \rangle, \nonumber \\ G_{\rm pair}(l) &=& \langle
\Delta_{j+l}^{\phantom\dagger}\,\Delta^{\dagger}_{j} \rangle,
\nonumber \\ \Delta_j &=& c_{j \, 2} \,c_{j \, 1}.
\end{eqnarray}
It is then immediate to obtain their Fourier transforms,
$n_{\sigma}(k)$ and $n_{\rm pair}(k)$. We emphasize that this
algorithm is exact: There are no approximations and the only errors
are statistical (which are in all cases smaller than the symbol size).
The polarization is given by $P=(N_2-N_1)/(N_2+N_1)$, where $N_1(N_2)$ is the
minority (majority) particle numbers.  The associated Fermi
wavevectors are $k_{\rm F \sigma} = 2 \pi N_\sigma/L$.

\begin{figure}[t]
\centerline{\epsfig{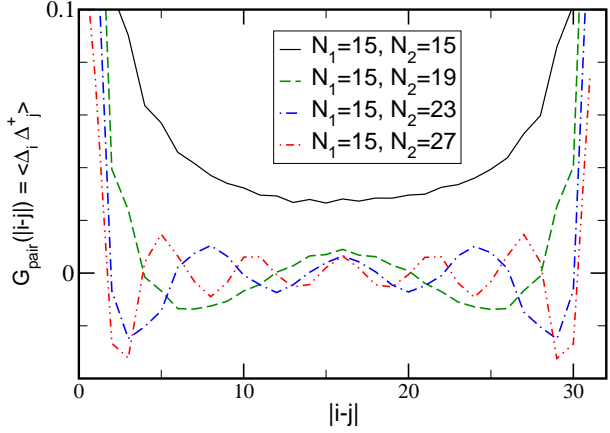}}
\caption{(color online).  The pair Green function $G_{\rm
pair}(|i-j|)$ for $N_1=15$ and $N_2=15, 19, 23, 27$ (Polarizations
$P=0, 0.118, 0.211$, and $0.286$) and $U=-8$.  For zero
polarization, $G_{\rm pair}(|i-j|)$ decays monotonically while for
$P\neq 0$, it oscillates, indicating the presence of an FFLO state.  }
\label{L32Um8Na15-PairGreenNb}
\end{figure}

We begin with the uniform system, $V_T=0$.  In one dimension, spin,
charge, and pair correlations of the Hubbard Hamiltonian in the ground
state decay algebraically with increasing separation, with smallest
exponent characterizing the phases \cite{g_ology}.  Here, with only an
attractive on-site interaction, the dominant order is in the $s$-wave
superconducting channel.  In Fig.~\ref{L32Um8Na15-PairGreenNb} we show
the real-space pair correlation function $G_{\rm pair}(l)$ for fixed
$U=-8$ and different $P$.  $G_{\rm pair}(l)$ decays monotonically for
$P=0$, but develops clear oscillations for $P \neq 0$.  These
oscillations are characteristic of the FFLO state: The mismatch of the
two Fermi surfaces leads to pairing at nonzero center of mass momentum
and, consequently, spatially inhomogenous regions in which the pairing
amplitude oscillates. The larger the Fermi surface mismatch, {\it
i.e.} the larger the $P$, the larger the center of mass momentum and
the smaller the period, as seen in
Fig.~\ref{L32Um8Na15-PairGreenNb}. The modulations in $G_{\rm
pair}(l)$ follow the Larkin-Ovchinnokov\cite{larkin64} (LO) form where
the order parameter is modulated with a ${\rm cos}(qr)$ as a function
of position $r$ with $q=\pm |k_{F1}-k_{F2}|$. The period, $T$, of the
oscillations is given by $T=2\pi/|q|$ as can be seen easily in
Fig.~\ref{L32Um8Na15-PairGreenNb}.

\begin{figure}[t]
\centerline{\epsfig{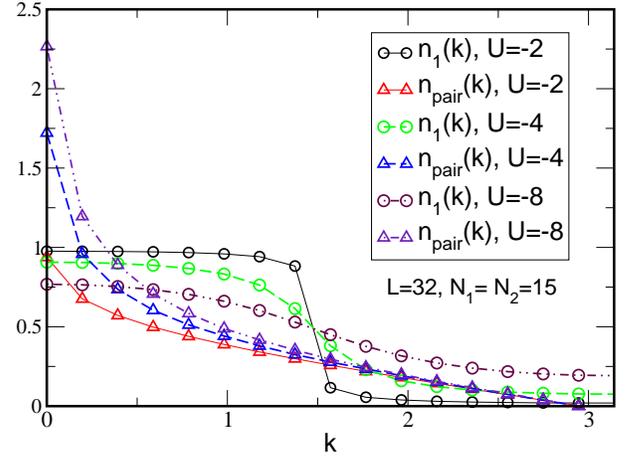}}
\caption{(color online).  $n_{1}(k)$ and $n_{\rm pair}(k)$ for the
case of $N_{1}=N_{2}=15$ and $L=32$, $\beta=64$, $U=-2,-4,-8$.  Here
$P=0$, so $n_1(k)=n_2(k)$.  }
\label{L32Na15Nb15-nk}
\end{figure}

To put the FFLO behavior in better context, we begin in
Fig.~\ref{L32Na15Nb15-nk} with $n_\sigma(k)$ and $n_{\rm pair}(k)$ for
the unpolarized case.  At weak coupling, $U=-2$, the fermion momentum
distribution function, $n_1(k)=n_2(k)$ has a sharp Fermi surface which
is then increasingly rounded as $|U|$ increases. In all cases,
$n_{\sigma}$ is a monotonic function of $k$.  Meanwhile the pair
momentum distribution function, $n_{\rm pair}(k)$, has a peak at $k=0$
which grows with $|U|$: The pairs have zero momentum and become more
tightly bound and numerous the larger the on-site attraction.

\begin{figure}[t]
\centerline{\epsfig{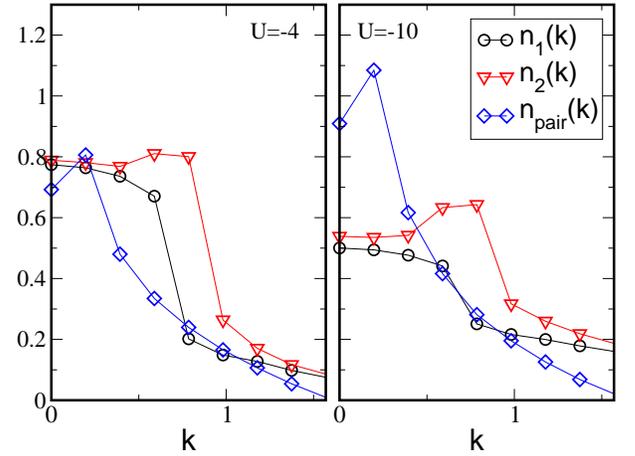}}
\caption{(color online).  $n_{1}(k)$, $n_2(k)$ and $n_{\rm pair}(k)$
for the case of $N_{1}=7$, $N_{2}=9$, $L=32$, $\beta=64$ and
$U=-4$ (left panel) and $U=-10$ (right panel).  }
\label{L32Na7Nb9-nk}
\end{figure}

When the system is polarized, it is thought to develop either pairs
with non-zero momentum via the FFLO mechanism, or zero momentum pairs
via the BP mechanism\cite{wilczek,koponen06,sarma}. We show in
Fig.~\ref{L32Na7Nb9-nk} the single particle and pair momentum
distributions for $N_1=7$, $N_2=9$ ($P=0.125$) with $U=-4$ and $U=-10$
on a system with $L=32$ sites. We note the following features: (a) in
both cases, $n_{\rm pair}(k)$ peaks at $\pm |k_{F1}-k_{F2}|$ (we show
only $k> 0$ since the figure is symmetric), (b) the height of the peak
increases with increasing $|U|$, (c) in both cases, the Fermi surfaces
for the minority and majority populations are much more sharply
defined than their $P=0$ counterparts at the same $|U|$.  Another
striking feature in the figure is the behavior of $n_2(k)$. For
$U=-4$, $n_2(k)$ displays a dip at $k=0.39<k_{F2}$ which deepens as
$|U|$ increases and spreads to $k=0$: For example, for $U=-10$, we see
that $n_2(k\leq 0.59)\approx 0.54$ but then rises to $n_2(0.59 < k\leq
0.78)\approx 0.64$ before it drops at the Fermi surface. This
remarkable feature, in which the momentum distribution can increase as
$k$ increases, is very robust for large negative $|U|$ and is not a
finite size effect.

Similar behavior has been seen in mean field both for FFLO and BP in
Ref.~\onlinecite{koponen06} where the Fulde-Ferrell (FF) single plane
wave {\it ansatz}, $\Delta(r) e^{iqr}$, was used. It was found that in
the BP case, $n_2(k)$ is symmetric around $k=0$ whereas in the FFLO
case $n_2(k)$ is asymmetric. It is likely that the LO {\it ansatz}
with cosine modulation due to the superposition of two plane waves
with opposite wavevectors will lead to a symmetric result for $n_2(k)$
in the FFLO state. However, there is no ambiguity in the pair momentum
distribution, $n_{\rm pair}(k)$: its peak lies at $k_{\rm peak}=\pm
|k_{F1}-k_{F2}|$ as is seen in Fig.~\ref{L32Na7Nb9-nk} where we only
display $k\geq 0$ since the figure is symmetric.  Furthermore, we note
that this peak starts to form at $k\neq0$ even for the smallest values
of $|U|$ we have examined. For example, for the case of
Fig.~\ref{L32Na7Nb9-nk}, it is present at $U=-0.5$.

\begin{figure}[t]
\centerline{\epsfig{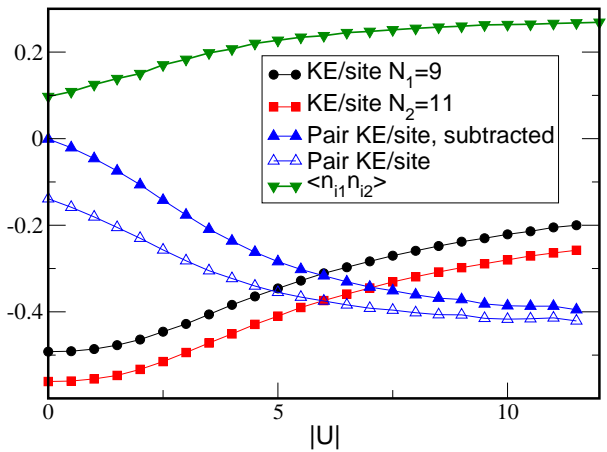}}
\caption{(color online). Kinetic energy/site for the minority and
  majority populations and for the pairs as functions of $|U|$ for
  $L=32$ and $\beta=64$. Also shown is the ``double occupancy"
  $\langle n_{i 1}n_{i 2}\rangle$ which gives the interaction energy
  when multiplied by $U$.}
\label{L32Na9Nb11-E}
\end{figure}

\begin{figure}[t]
\centerline{\epsfig{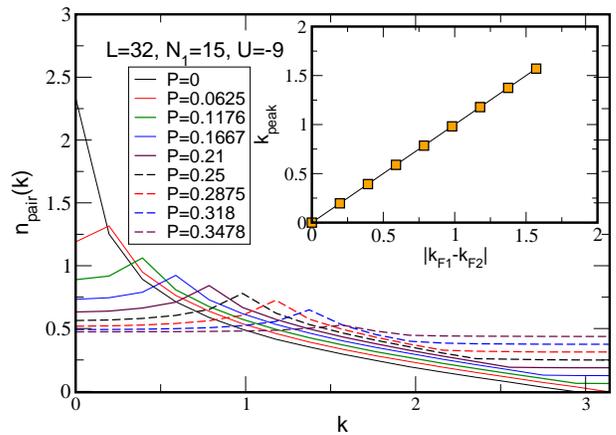}}
\caption{(color online).  $n_{\rm pair}(k)$ for several values of the
polarization showing the FFLO peak going to higher values of $k_{\rm
peak}$ as $|k_{F1}-k_{F2}|$ increases. The inset shows the position of
the peak versus $|k_{F1}-k_{F2}|$.  }
\label{L32Na15Um9-kpeakvsk}
\end{figure}

Additional insight on the pairing is obtained from the energetics. In
Fig.~\ref{L32Na9Nb11-E} we show the kinetic energy (KE) per site for
the minority and majority populations, $\langle
c_{l+1\,\sigma}^\dagger c_{l \sigma}^{\phantom\dagger} \rangle $, and
for the pairs, $\langle \Delta_{l+1}^{\phantom\dagger}
\Delta^{\dagger}_{l}\rangle$. We also show the subtracted pair KE/site
($\langle \Delta_{l+1}^{\phantom\dagger} \Delta^{\dagger}_{l} \rangle
-\langle c_{l\, 1}^\dagger c_{l\,1}^{\phantom\dagger} \rangle \langle
c_{l\,2}^\dagger c_{l\, 2}^{\phantom\dagger} \rangle $) to emphasize
the contribution from pairing. We see that as $|U|$ increases, the
single fermion kinetic energies decrease (in absolute value) while the
pair KE increases: more and more of the fermion hopping is performed
in pairs. We also see clear cross-over behavior in the KE curves
(change in the sign of the curvature) as $|U|$ is increased which can
be understood by examining the double occupancy, $\langle n_{i 1}n_{i
2}\rangle$. Clearly, $\langle n_{i 1}n_{i 2}\rangle$ lies between
$N_1N_2/L^2$ at $U=0$ (no pairing) and $N_1/L$ at very large negative
$U$ (maximum pairing). When pairing is saturated, the system is made
of $N_1$ tightly bound pairs and a gas of $N_2-N_1$ unpaired
fermions. The crossover in the KE appears to take place as the number
of tightly bound pairs most rapidly approaches its saturation value at $U
\approx -2.5$. We emphasize that for {\it all} values of $U$, the peak
in $n_{\rm pair}$ is at $k_{\rm peak}\neq 0$ when $P\neq 0$.

To underscore the robustness of FFLO pairing, we show $n_{\rm
pair}(k)$ in Fig.~\ref{L32Na15Um9-kpeakvsk} at $U=-9$ for several
polarizations.  The curves clearly show a peak at $k_{\rm peak}\neq 0$
which, as is seen in the inset, corresponds to $|k_{F1}-k_{F2}|$. It
is striking that even at the largest polarizations considered,
corresponding to $N_1=15$ and $N_2=31$, FFLO pairing is still present
as seen in the $n_{\rm pair}(k)$ peak. For these parameters, we have
not found the Clogston-Chandrasekhar limit\cite{clog} in which extreme
polarization completely destroys the superconducting state. It is
possible that going to much lower $N_1$ can reach it.


Several measurements have been reported on spin imbalanced cold atom
systems \cite{zwierlein06,partridge06} confined in 3-dimensional, but
highly elongated, traps.  In these experiments, the effect of the
confining potential is critical to the results.  We now apply a trap,
$V_T\neq 0$, in our simulations and discuss its effect on the FFLO
pairing state. It is clear that the presence of the majority species
in the trap center, combined with the attractive interaction, will
lead to increased localization of the minority species in the center.
Experiments\cite{zwierlein06,partridge06} show additional interesting
features: this localization tendency is so marked that the local
density difference $n_{i 1}-n_{i 2}$ vanishes in an extended region
about the trap center.  One recent DMRG calculation\cite{tezuka07},
observes both the FFLO state and the squeezing of the minority
population, but their local density difference is {\it maximal} at the
trap center.  More precisely, there is an extended flat region of
constant local polarization near the center, which then falls off as
the distance increases.  The constancy of this polarization is
reflected in the fact that the period of the FFLO oscillations is
uniform throughout the central region.

\begin{figure}[t]
\centerline{\epsfig{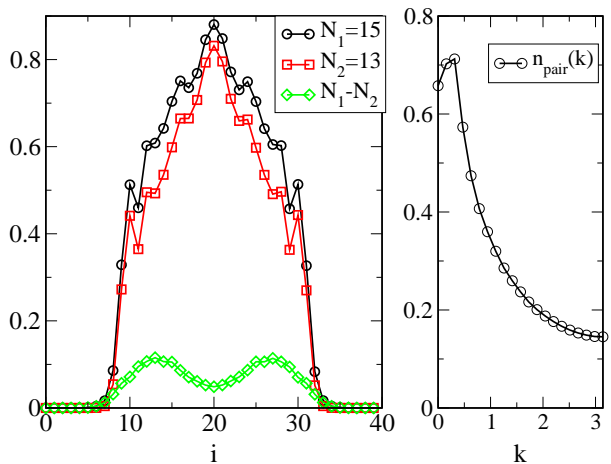}}
\caption{(color online).  Left: Density profiles for the minority and
    majority populations in a trap $V_T=0.005$.  Here $L=40, U=-10,
    \beta=64$.  The phase in the trap center is not a fully packed
    Fock state, yet the difference in the local density still exhibits
    a deep minimum.  Right: The Fourier transform $n_{\rm pair}(k)$ of
    the pair correlation function.  The peak position is at $k=\pi
    /10$, which is very close to the value $k=2 \pi (n_1-n_2) = 0.31$
    obtained by using the the local polarization $p_i=n_{i 1}-n_{i
    2}=0.049$ at the trap center $i=20$.  }
\label{denprof-trap}
\end{figure}

A second DMRG treatment\cite{feiguin07} reports zero density
difference in the trap center associated with a Fock state of sites
which are fully occupied with $n_{i 1}=n_{i 2}=1$.  This leads to a
vanishing local polarization, but does not support superconducivity
(at any wavevector) since the particles cannot move.  At larger global
polarization the Fock state melts and the minimum in local
polarization is nonvanishing at the trap center, but the state there
is metallic rather than superconducting, since the sites are still
fully occupied in the majority channel.

In Fig.~\ref{denprof-trap} we show that a marked minimum in the
density difference is present even when the trap center is superfluid,
with both 
$n_{i 1} < 1$ and $n_{i 2} < 1$, as in the experimental situation.  The
figure also shows a peak in $n_{\rm pair}(k)$ at nonzero $k$,
demonstrating that pairing occurs and, furthermore, that this
superfluid is of the FFLO variety, as expected from the nonzero local
polarization, $p_i$.  At the trap center, $p_i=n_{i 1}-n_{i 2} =
0.049$ which would lead to FFLO pairing at $k_a= 0.31$ in the uniform
case.  Towards the edge of the central polarization minimum, the
maximal $p_i=0.115$, with an associated $k_b= 0.72$.  The observed
peak in $n_{\rm pair}(k)$ lies very close to $k_a$ and is determined
by the lower polarization region.


In conclusion, we have studied the one-dimensional attractive Hubbard
model with imbalanced populations for a range of $P$ and $U$ values
with and without a trapping potential. In the absence of a trap, we
have presented results for pair Green functions, single particle and
pair momentum distributions and kinetic and interaction energies. Our
results show that FFLO pairing is surprisingly robust from very small
values of $|U|$ all the way up to saturation at very large $|U|$ and
for a wide range of $P$. We have found no values for $U$ and $P$ for
which BP dominates over FFLO: the maximum of $n_{\rm pair}(k)$ is
always at $k_{\rm peak}=\pm |k_{F1}-k_{F2}|$.  We have also shown that
at large $|U|$, $n_{\sigma}(k)$ of the majority populations is not
monotonic. Finally we have shown that these features are robust to the
spatial inhomogeneities induced by the presence of a trap.

GGB is supported by the CNRS (France) PICS 18796, RTS by DARPA/AFOSR,
MHH by NSF PHY-0649297 (REU) and VGR by the research program of the
`Stichting voor Fundamenteel Onderzoek der Materie (FOM)'. We
acknowledge useful discussions with A. Muramatasu and G.~Jules.

\end{document}